%% file: main.tex
\documentclass[9pt,conference]{IEEEtran}
\IEEEoverridecommandlockouts

\usepackage{cite}
\usepackage{amsmath,amssymb,amsfonts}
\usepackage{algorithmic}
\usepackage{graphicx}
\usepackage{textcomp}
\usepackage{xcolor}

\usepackage{array}
\usepackage{mathtools}
\usepackage{tcolorbox}
\usepackage{color}
\usepackage{theorem}
\usepackage{amssymb}
\usepackage{caption}
\usepackage{subcaption}
\usepackage{cite,hyperref}
\usepackage{cases}
\usepackage{url}
\usepackage{algorithmic}
\usepackage{algorithm}
\usepackage{tikz}
\usetikzlibrary{shapes,arrows}
\usepackage{enumitem}

\usepackage{stmaryrd}

\usepackage{multirow}
\usepackage{hhline}

\usepackage{tikz}
\usepackage{moresize}
\usepackage{pgfplots}
\usepackage{wrapfig}
\pgfplotsset{compat=1.17}
\pgfplotstableset{col sep=comma}
\newtcolorbox{myblockt}[1]{colback=urblue!5!white,
	colframe=urblue,fonttitle=\bfseries,
	title=#1}
\newtcolorbox{myblock}{colback=urblue!5!white,
	colframe=urblue,fonttitle=\bfseries}

\def\BibTeX{{\rm B\kern-.05em{\sc i\kern-.025em b}\kern-.08em
    T\kern-.1667em\lower.7ex\hbox{E}\kern-.125emX}}

\input{mysymbol.sty}

\pgfplotsset{compat=1.17}
\pgfplotstableset{col sep=comma}
\tikzset{every mark/.append style={scale=1.5, solid}, font=\footnotesize}
\pgfplotsset{
    width=1.05\textwidth,
    legend style={
        font=\ssmall ,  
        inner xsep=1pt,
        inner ysep=1pt,
        nodes={inner sep=1pt}},
    legend cell align=left,
	every axis/.append style={line width=0.5pt},
	every axis plot/.append style={line width=1.25pt},
    every axis y label/.append style={yshift=-5pt}
}

\begin{document}

\title{Low-Rank Tensors for Multi-Dimensional Markov Models
\thanks{This work was partially supported by the NSF under award CCF-2340481, the Spanish AEI PID2022-136887NB-I00 and the Community of Madrid via the ELLIS Madrid Unit. Research was sponsored by the Army Research Office and was accomplished under Grant Number W911NF-17-S-0002. The views and conclusions contained in this document are those of the authors and should  not be interpreted as representing the official policies, either expressed or implied, of the Army Research Office or the U.S. Army or the U.S. Government. The U.S. Government is authorized to reproduce and distribute reprints for Government purposes notwithstanding any copyright notation herein.
Emails:  \href{mailto:nav@rice.edu}{nav@rice.edu}, \href{mailto:s.rozada.2019@alumnos.urjc.es}{s.rozada.2019@alumnos.urjc.es}, \href{mailto:antonio.garcia.marques@urjc.es}{antonio.garcia.marques@urjc.es}, \href{mailto:segarra@rice.edu}{segarra@rice.edu}
}
}

\author{
\IEEEauthorblockN{
Madeline Navarro\IEEEauthorrefmark{1},
Sergio Rozada\IEEEauthorrefmark{2},
Antonio G. Marques\IEEEauthorrefmark{2},
Santiago Segarra\IEEEauthorrefmark{1}
} %
\IEEEauthorblockA{
\IEEEauthorrefmark{1}Dept. of Electrical and Computer Engineering, Rice University, Houston, TX, USA } %
\IEEEauthorblockA{
\IEEEauthorrefmark{2}Dept. of Signal Theory and Communications, Rey Juan Carlos University, Madrid, Spain } %
}

\maketitle

\begin{abstract}
    This work presents a \textit{low-rank tensor} model for \textit{multi-dimensional Markov chains}.
    A common approach to simplify the dynamical behavior of a Markov chain is to impose low-rankness on the transition probability matrix.
    Inspired by the success of these matrix techniques, we present low-rank tensors for representing transition probabilities on multi-dimensional state spaces.
    Through tensor decomposition, we provide a connection between our method and classical probabilistic models.
    Moreover, our proposed model yields a parsimonious representation with fewer parameters than matrix-based approaches.
    Unlike these methods, which impose low-rankness uniformly across all states, our tensor method accounts for the multi-dimensionality of the state space.
    We also propose an optimization-based approach to estimate a Markov model as a low-rank tensor.
    Our optimization problem can be solved by the alternating direction method of multipliers (ADMM), which enjoys convergence to a stationary solution.
    We empirically demonstrate that our tensor model estimates Markov chains more efficiently than conventional techniques, requiring both fewer samples and parameters.
    We perform numerical simulations for both a synthetic low-rank Markov chain and a real-world example with New York City taxi data, showcasing the advantages of multi-dimensionality for modeling state spaces.
\end{abstract}


\section{Introduction}
\label{S:intro}

Dynamic processes provide a mathematical framework for modeling the evolution of complex systems~\cite{birkhoff1927dynamical, katok1995introduction}. 
When the system and its dynamics are known, control theory has a long history of successful applications \cite{glasser1985control, glad2018control}. 
However, the dynamics are not always available and oftentimes probabilistic, and they must instead be estimated from sampled trajectories.
Learning the dynamics of complex systems is challenging due to non-linear behavior, the presence of \textit{uncertainties}, high dimensionality, and data scarcity~\cite{brunton2022data}.
Assuming specific properties of the system can render the problem tractable.
In this work, we rely on two assumptions: Markovianity, where future behavior relies only on the current state of the system, and low-rankness~\cite{chi2019nonconvex}, where the system dynamics can be represented by a small set of hidden variables~\cite{thibeault2024low}.

Low-rank methods are widely used for learning the transition probabilities of Markov systems, which can be represented via probability transition matrices.
When the full matrix is known, low-rank techniques enable compression for lower-dimensional models~\cite{deng2011optimal,deng2012model}.
Moreover, low-rankness has shown success in estimating an unknown probability matrix from samples, a classical problem in estimation theory~\cite{lehmann2006theory,han2015minimax}.
Markov chains pose a challenging case, where samples within a trajectory of observed states are typically correlated.
Many works propose low-rank Markov transition matrix estimation for a simpler problem that is feasible for limited samples~\cite{li2018estimation,zhang2019spectral,zhu2022learning}.
Beyond practical considerations, the concept of low-rankness arises naturally when transitions occur in a low-dimensional hidden space, such as for hidden Markov chains~\cite{hsu2012spectral,huang2018recovering,fraser2008hidden}.

However, past literature does not fully exploit the ubiquitous multi-dimensionality of state spaces. 
Tensors provide natural models for transition probabilities between multi-dimensional states, where tensor dimensions correspond to those of the state space~\cite{ding2020multiuser,kuang2016tensorbased,wang2020m2t2}. 
By considering the tensor to be low-rank, we postulate a parsimonious model that also deals with the curse of dimensionality.
We thus leverage the notion of low-rankness for tensors to formulate an interpretable model of the transition probabilities, both for the entire state space and for each dimension.
Previous works exploit tensor low-rankness for learning Markov models via higher-order moments~\cite{anandkumar2014tensor,azizzadenesheli2016reinforcement,li2014limiting}, for locally-interacting Markov chains~\cite{kressner2014low,georg2023low,buchholz2000multilevel,kressner2014lowrank}, and for value functions of multi-dimensional Markov decision processes~\cite{tsai2021tensor, rozada2024tensor,dolgov2021tensor,gorodetsky2018highdimensional,kuinchtner2021cpmdpa}.
However, this is the first work to use low-rank tensors for modeling transitions in such spaces.
Moreover, we consider the practical task of estimating a transition tensor from trajectories via a provably convergent optimization approach.
Our contributions are as follows.
\begin{itemize}[left= 5pt .. 15pt, noitemsep]
    \item[\textbf{C1}]
    We propose a tensor-based probabilistic model for Markov chains on multi-dimensional state spaces, which provides a link between probabilistic models and low-rank tensor decompositions.
    \item[\textbf{C2}]
    We present an optimization-based approach to estimate a low-rank Markov transition tensor from trajectories, which enjoys convergence when solved with alternating minimization.
    \item[\textbf{C3}]
    We demonstrate the advantages of tensors for modeling Markovian systems and the viability of their estimation through synthetic and real-world simulations.
\end{itemize}

\section{Tensor low-rank model}
\label{S:model}

In this section, we first relate tensors to Markov models.
Then, we introduce our probabilistic tensor model, for which we describe the consequences of tensor low-rankness.

\subsection{Preliminaries}
\label{Ss:prelim}
We consider an ergodic Markov chain with a finite $D$-dimensional state space $\ccalS := \ccalS_1 \times \cdots \times \ccalS_D$, where $\ccalS_d := [I_d] = \{1, 2, \dots, I_d\}$ for every $d\in[D]$. 
For each $\bbs,\bbs'\in\ccalS$, the transition probability $p(\bbs'|\bbs)$ is the conditional probability of transitioning to state $\bbs'$ from state $\bbs$.
Learning a Markov model is equivalent to estimating the transition distribution $p$.
Since we assume ergodicity, the Markov model has a unique stationary distribution $\pi$, where $\pi(\bbs)$ is the probability of being in the state $\bbs\in\ccalS$.
Additionally, let the joint probability distribution be $q$, where $q(\bbs, \bbs') = p(\bbs'|\bbs) \pi(\bbs)$ models the likelihood of the co-occurrence of two consecutive states $\bbs, \bbs' \in \ccalS$.
We exploit our knowledge of the multi-dimensional state space by arranging the probabilities $p(\bbs'|\bbs)$ in a tensor format.
Specifically, consider the state $\bbs \in \ccalS$, which is a vector $\bbs = [s_1, \dots, s_D]^T$ for $s_d\in\ccalS_d$.
We can thus arrange the transition probabilities $p(s'_1, \dots, s'_D | s_1, \dots, s_D)$ into the transition probability tensor $\tenbP \in [0, 1]^{I_1 \times \dots \times I_D \times I_1 \times \dots \times I_D}$, where $\tenbP(\bbs, \bbs') = p(\bbs'|\bbs)$ for every $\bbs,\bbs'\in\ccalS$.
Analogously, we also define the stationary probability tensor $\tenbR\in [0, 1]^{I_1 \times \dots \times I_D}$ for $\pi$ and the joint probability tensor $\tenbQ\in [0, 1]^{I_1 \times \dots \times I_D \times I_1 \times \dots \times I_D}$ for $q$.

\subsection{Probabilistic Model}
\label{Ss:model}
Similar to existing works that learn Markov chains via low-rank decomposition~\cite{zhang2019spectral}, we propose first estimating the joint distribution $q$, from which we derive the transition distribution $p$.
This enables a connection between probabilistic models and low-rank tensor decompositions, as we will show. 
In particular, we model the joint probability tensor $\tenbQ$ with a low-rank tensor decomposition.
To derive the decomposition of $\tenbQ$, we first note that any joint distribution can be represented using a naive Bayes model~\cite{koller2009probabilistic}, which assumes the existence of a hidden dimension with $F$ possible values that generates $q$~\cite[Proposition 1]{kargas2018tensors}.
Formally, for every $\bbs,\bbs'\in\ccalS$ we model $q$ as
\begin{equation}
    \label{eq::naive_bayes}
    q(\bbs, \bbs') = \sum_{f=1}^F \text{Pr}(f) \prod_{d=1}^D \text{Pr}(s_d | f) \prod_{d=1}^D \text{Pr}(s'_d | f),
\end{equation}
which can be interpreted as a tensor decomposition with rank $F$~\cite{shashua2005non,lim2009nonnegative,kargas2018tensors}.
To see this, first assume that $\tenbQ$ has rank $F$, which means that it can be represented by the sum of at minimum $F$ rank-1 tensors. 
This sum, known as the canonical polyadic decomposition (CPD)~\cite{sidiropoulos2017tensor}, is defined as $\tenbQ = \sum_{f=1}^{F} \bblambda(f) \bbQ_1(:, f) \circ \cdots \circ \bbQ_D(:, f) \circ \bbQ'_1(:, f) \circ \cdots \circ \bbQ'_{D}(:, f)$, where $\circ$ is the outer product, $\bbQ_d, \bbQ'_d \in \mathbb{R}^{I_d \times F}$ are known as factor matrices, and $\bblambda$ is a normalization term.
We denote this decomposition as $\tenbQ = [[\bblambda, \bbQ, \bbQ']]$, with $\bbQ:=\{\bbQ_d\}_{d=1}^D$ and $\bbQ':=\{\bbQ'_d\}_{d=1}^D$. 
Then, under the CPD model, the entry of $\tenbQ$ associated with the states $\bbs, \bbs' \in \ccalS$ is computed as
\begin{equation}
    \label{eq::cpd_model}
    \tenbQ(\bbs, \bbs') = \sum_{f=1}^F \bblambda(f) \prod_{d=1}^D \bbQ_d(s_d | f) \prod_{d=1}^D  \bbQ'_d(s'_d | f),
\end{equation}
which is equivalent to~\eqref{eq::naive_bayes} substituting our tensor formulation of $\tenbQ$ and its decomposition $\cp{\bblambda,\bbQ,\bbQ'}$, that is, $\bbQ_d(s_d | f) = \text{Pr}(s_d | f)$, $ \bbQ'_d(s'_d | f) = \text{Pr}(s'_d | f)$, and $\bblambda(f) = \text{Pr}(f)$.
Thus, estimating the CPD decomposition of $\tenbQ$ is equivalent to estimating $\bblambda$ as the distribution over the hidden variable and the columns of $\bbQ_d$ and $\bbQ_d'$ as distributions over the dimensions of the state space.
Critically, under mild dimensionality conditions the decomposition in~\eqref{eq::cpd_model} and thus the naive Bayes model in~\eqref{eq::naive_bayes} are unique~\cite{sidiropoulos2017tensor}, and so are the distributions for the latent and state dimensions $\bblambda$, $\bbQ_d$, and $\bbQ_d'$.
However, there may exist other naive Bayes representations with more than $F$ hidden states that generate the same distribution $q$.

Next, we obtain a tensor representation $\tenbR$ for the stationary probability distribution $\pi$ from the CPD model for $\tenbQ$.
To do so, we compute the marginal distribution $\pi(\bbs) = \sum_{\bbs' \in \ccalS} q(\bbs, \bbs')$ as
\begin{align}
    \notag
    \tenbR(\bbs)=\sum_{\bbs' \in \ccalS} \tenbQ(\bbs, \bbs') &= \sum_{f=1}^F \bblambda(f) \prod_{d=1}^D \bbQ_d(s_d | f) \sum_{\bbs' \in \ccalS} \prod_{d=1}^D \bbQ'_d(s'_d | f) \\
    \label{eq::cpd_marginal}
    &= \sum_{f=1}^F \bblambda(f) \prod_{d=1}^D \bbQ_d(s_d | f).
\end{align}
Note that the decomposition of $\tenbR$ in \eqref{eq::cpd_marginal} may not be the CPD decomposition of $\tenbR$ if the tensor rank of $\tenbR$ is lower than $F$.
Moreover, by ergodicity we may also obtain $\tenbR$ by marginalizing over $\bbs$ instead of $\bbs'$ since $\pi(\bbs) = \sum_{\bbtau\in\ccalS} q(\bbs,\bbtau) = \sum_{\bbtau\in\ccalS} q(\bbtau,\bbs)$~\cite{norris1998markov}.
Thus, we may exchange $\bbQ'_d$ for $\bbQ_d$ in the second line of~\eqref{eq::cpd_marginal}.

Lastly, we derive the transition probability tensor $\tenbP$. 
We apply the chain rule of probability with $\tenbQ$ and $\tenbR$ to obtain
\begin{equation}
    \label{eq::cpd_joint}
    \tenbP(\bbs, \bbs') = \frac{\tenbQ(\bbs, \bbs')}{\tenbR(\bbs)}.
\end{equation}
Recall that the primary objective in learning Markov chains is to estimate $\tenbP$. 
Given $\tenbQ$, we can compute $\tenbR$ using~\eqref{eq::cpd_marginal} then $\tenbP$ using~\eqref{eq::cpd_joint}.
In practice, however, we do not have direct access to $\tenbQ$. 
The next section addresses the challenge of estimating the joint probability tensor $\tenbQ$ from observed trajectories sampled from the chain.


\begin{figure*}[t]
    \centering
    \begin{minipage}[c]{.19\textwidth}
        \includegraphics[width=\textwidth]{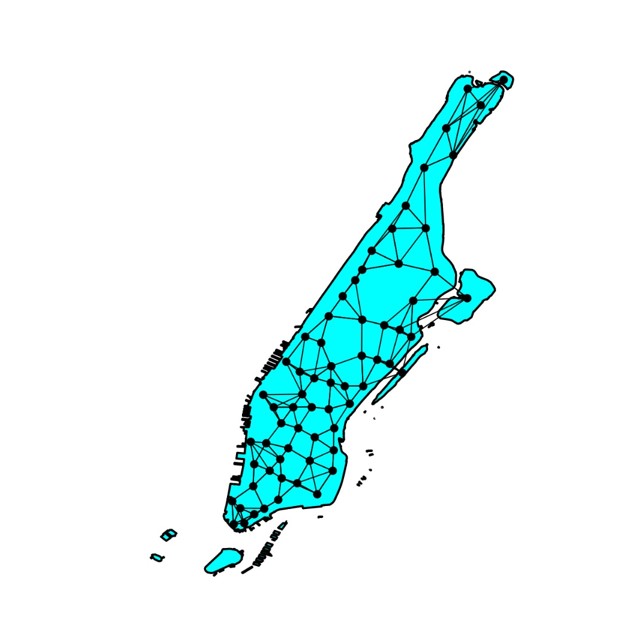}
        \centering{\footnotesize (a) NYC taxi map}
    \end{minipage}
    \hspace{0.3cm}
    \begin{minipage}[c]{.19\textwidth}
        \includegraphics[width=\textwidth]{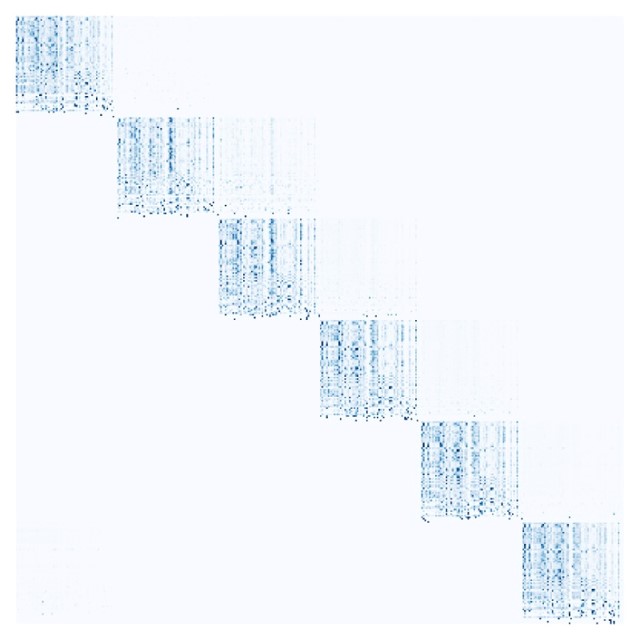}
        \centering{\footnotesize (b) True transition tensor $\tenbP$}
    \end{minipage}
    \hspace{0.3cm}
    \begin{minipage}[c]{.19\textwidth}
        \includegraphics[width=\textwidth]{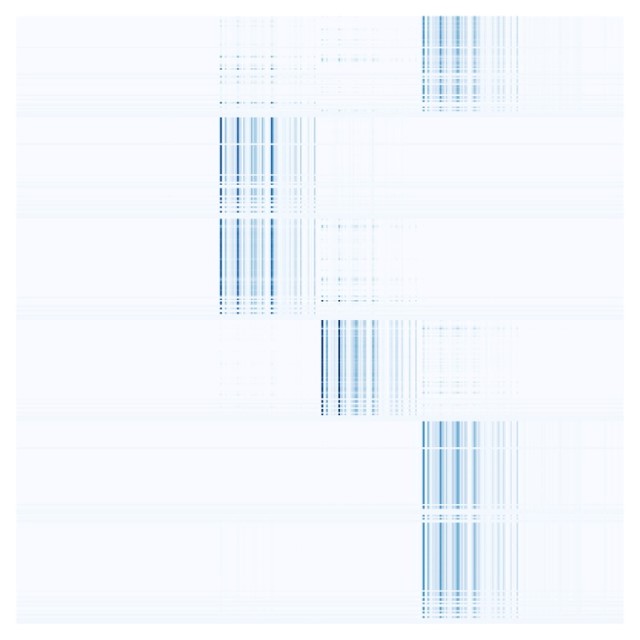}
        \centering{\footnotesize (c) \textbf{SLRM} estimated tensor}
    \end{minipage}
    \hspace{0.3cm}
    \begin{minipage}[c]{.19\textwidth}
        \includegraphics[width=\textwidth]{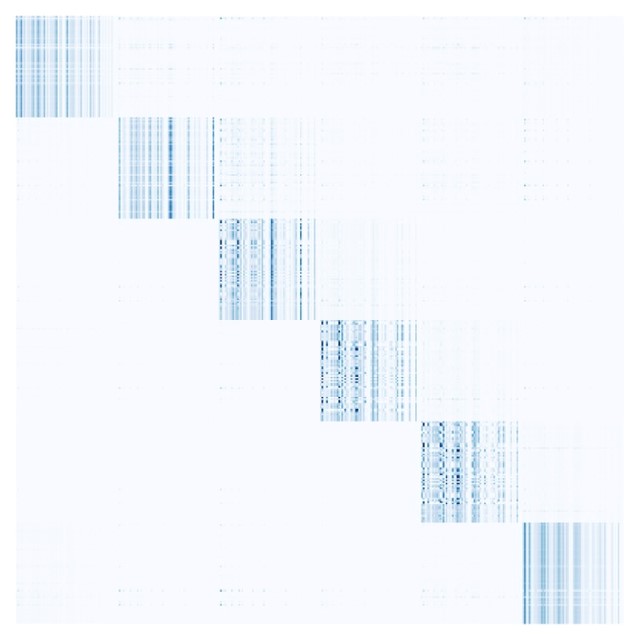}
        \centering{\footnotesize (d) \textbf{LRT} estimated tensor}
    \end{minipage}
    \caption{
    Visualizations of a real-world Markov chain for NYC taxi data.
    (a) 
    NYC map of taxi pick-up and drop-off locations.
    The graph indicates the state space, where nodes correspond to states and edges show a subset of possible transitions.
    (b)
    True transition tensor $\tenbP$ for NYC taxi Markov model. 
    (c)
    Estimated transition tensor $\hat \tenbP$ from matrix method \textbf{SLRM}~\cite{zhang2019spectral}.
    (d)
    Estimated transition tensor $\hat \tenbP$ from proposed tensor method \textbf{LRT}.
    }
    \vspace{-.3cm}
\label{f:ny_data}
\end{figure*}


\section{Low-rank Probability Tensor Estimation}
\label{S:optim}

Given our proposed tensor model, we now aim to learn the transition probabilities in $\tenbP$ from a trajectory of length $N$ of multi-dimensional states $\ccalX = \{\bbs_n\}_{n=1}^N$ observed on $\ccalS$.
To this end, we exploit the naive Bayes representation of $q$ by estimating the CPD decomposition of $\tenbQ$ in  \eqref{eq::cpd_model}, from which we obtain estimates for $\tenbR$ via~\eqref{eq::cpd_marginal} and $\tenbP$ via~\eqref{eq::cpd_joint}.
First, let $\tilde{\tenbQ}$ be the empirical estimate of the joint distribution from $\ccalX$~\cite{hao2018learning}, where for every $\bbs,\bbs'\in\ccalS$,
\begin{equation}\label{eq::emp_est_Q}
    \tilde{\tenbQ}(\bbs,\bbs')
    =
    \frac{1}{N}
    \left|\left\{ n\in[N-1]~|~ \bbs_n = \bbs, \bbs_{n+1}=\bbs' \right\}\right|.
\end{equation}
Thus, each entry $\tilde{\tenbQ}(\bbs,\bbs')$ denotes the empirical frequency of the co-occurrence of consecutive states $\bbs$ and $\bbs'$.
Note that we may also estimate the transition distribution $\tilde{\tenbP}$, where for every $\bbs,\bbs'\in\ccalS$ such that $\bbs$ is observed at least once, that is, $\bbs_n=\bbs$ for some $n\in[N]$,
\begin{equation}\label{eq::emp_est_P}
    \tilde{\tenbP}(\bbs,\bbs') = 
    \frac{ N \tilde{\tenbQ}(\bbs,\bbs') }{ \left|\left\{ n\in[N-1]~|~ \bbs_n = \bbs \right\}\right| },
\end{equation}
otherwise, if $\bbs$ is never observed, then we let $\tilde{\tenbP}(\bbs,\bbs') = I^{-1}$ for $I := \prod_{d=1}^D I_d$.
With $\tilde{\tenbQ}$, we propose the following optimization problem to estimate the decomposition $\cp{\bblambda,\bbQ,\bbQ'}$ for $\bblambda\in[0,1]^F$, $\bbQ_d,\bbQ'_d\in[0,1]^{I_d\times F}$ as our rank-$F$ estimate of $\tenbQ$
\alna{
    & \min_{\bblambda, \bbQ, \bbQ'} ~& 
    g(\bblambda,\bbQ, \bbQ') :=
    \frac{1}{2} \norm{ \tilde{\tenbQ} - \cp{\bblambda,\bbQ, \bbQ'} }_F^2
&\label{eq::opt_prob}\\&
    &\mathrm{s.t.}~~~~~& \!\!\!\!
    \bblambda, \bbQ, \bbQ' \geq \bbzero, ~ 
    \bbone^\top \bblambda = 1, ~
    \bbQ_d^\top \bbone = (\bbQ'_d)^\top \bbone = \bbone, ~
    \forall d\in[D],
\nonumber
}
with constraints to ensure valid distributions of $\bblambda$, $\bbQ_d$, and $\bbQ_d'$.
In particular, the first constraint imposes element-wise non-negativity, while the equality constraints ensure that $\bblambda$ and each of the columns of $\bbQ_d$ and $\bbQ_d'$ sum to 1.
From~\eqref{eq::opt_prob}, we obtain an estimate $\hat{\tenbQ}$, which we use to obtain $\hat{\tenbR}$ using~\eqref{eq::cpd_marginal}. 
Then, we use~\eqref{eq::cpd_joint} to estimate $\hat{\tenbP}$.

Alternating minimization methods are popular choices for obtaining tensor decompositions~\cite{fu2020computing}.
Indeed, while the objective $g$ in~\eqref{eq::opt_prob} is non-convex, it is convex in each element of the decomposition, $\bblambda$, $\bbQ_d$, and $\bbQ_d'$ for every $d\in[D]$.
The problem~\eqref{eq::opt_prob} is then amenable to the alternating direction method of multipliers (ADMM), a successful approach for tensor decomposition with additional assumptions~\cite{huang2016flexible, kargas2018tensors}.
We thus obtain $\hblambda$, $\hbQ$, and $\hbQ'$ via ADMM, which yields our estimate $\hat{\tenbQ}$, from which we derive $\hat{\tenbR}$ and then $\hat{\tenbP}$.

To apply ADMM, we first introduce auxiliary variables $\bbpsi$, $\bbS = \{\bbS_d\}_{d=1}^D$, and $\bbS'=\{\bbS'_d\}_{d=1}^D$, which act as copies of $\bblambda$, $\bbQ$, and $\bbQ'$, respectively, and present an equivalent problem
\begin{subequations}
\alna{
    & \min_{\bblambda, \bbQ, \bbQ', \bbpsi, \bbS, \bbS'} & ~~~
    g(\bblambda,\bbQ, \bbQ') 
    + \mbI\{ \bblambda\geq \bbzero, \bbQ\geq \bbzero,\bbQ' \geq \bbzero \}
&\label{eq::aux_prob}\\&
    &\mathrm{s.t.} \qquad& ~~~
    \bblambda=\bbpsi,~
    \bbQ_d=\bbS_d,~
    \bbQ_d'=\bbS_d',~
& \label{eq:const_admm_aux}\\&
    &&~~~
    \bbone^\top \bbpsi = 1, ~
    \bbS_d^\top \bbone = (\bbS'_d)^\top \bbone = \bbone, ~
    \forall d\in[D],\label{eq:const_admm_sumone}
}
\end{subequations}
where $\mbI\{\cdot\}$ denotes an indicator function that is 0 when its argument is true and infinity otherwise. We introduce the dual variables $\bbu$, $\bbU = \{\bbU_d\}_{d=1}^D$, $\bbU' = \{\bbU'_d\}_{d=1}^D$, which are associated with the constraints in \eqref{eq:const_admm_aux}, and $v$, $\bbv=\{\bbv_d\}_{d=1}^D$, $\bbv' = \{\bbv'_d\}_{d=1}^D$, associated with \eqref{eq:const_admm_sumone}. For convenience, we collect the primal variables in $\bbX := (\bblambda,\bbQ,\bbQ')$, the auxiliary variables in $\bbY := (\bbpsi,\bbS,\bbS')$, and the dual variables in $\bbW := (\bbu,\bbU,\bbU',v,\bbv,\bbv')$.
The augmented Lagrangian for~\eqref{eq::aux_prob} is
\alna{
    \ccalL_{\beta}(\bbX,\bbY,\bbW)
    =
    g(\bblambda,\bbQ,\bbQ')
    + \mbI\{ \bblambda\geq \bbzero, \bbQ\geq \bbzero,\bbQ' \geq \bbzero \}
&\label{eq::auglag}\\&
    \quad
    + \frac{\beta}{2}
    \sum_{d=1}^D
    \left(
    \norm{ \bbQ_d-\bbS_d }_F^2
    + 
    \norm{ \bbQ_d'-\bbS_d' }_F^2
    \right)
    + \frac{\beta}{2}
    \norm{\bblambda-\bbpsi}_2^2
&\nonumber\\&
    \quad
    + \frac{\beta}{2}
    \sum_{d=1}^D 
    \left(
    \norm{ \bbS_d^\top\bbone - \bbone }_F^2
    + 
    \norm{ (\bbS_d')^\top\bbone - \bbone }_F^2
    \right)
    + \frac{\beta}{2}
    (\bbone^\top\bbpsi - 1)^2
&\nonumber\\&
    \quad
    +
    \sum_{d=1}^D 
    \Big(
        \inner{ \bbU_d, \bbQ_d-\bbQ_d }
        +
        \inner{ \bbU_d', \bbQ_d'-\bbQ_d' }
    \Big)
    +
    \inner{ \bbu, \bblambda-\bbpsi }
&\nonumber\\&
    \quad
    +
    \sum_{d=1}^D 
    \Big(
        \inner{ \bbv_d, \bbS_d^\top\bbone - \bbone }
        +
        \inner{ \bbv_d', (\bbS_d')^\top\bbone - \bbone }
    \Big)
    +
    v (\bbone^\top\bbpsi - 1)
\nonumber
}
for ADMM parameter $\beta > 0$.
The process for ADMM minimizes $\ccalL_{\beta}$ for each variable cyclically, that is, we first optimize $\ccalL_{\beta}$ for each primal variable $\bblambda$, $\bbQ_d$, $\bbQ'_d$ for every $d\in[D]$ then each auxiliary variable $\bbpsi$, $\bbS_d$, $\bbS'_d$ for every $d\in[D]$, followed by the dual variable updates $\bbu$, $\bbU_d$, $\bbU'_d$, $v$, $\bbv_d$, $\bbv'_d$ for every $d\in[D]$~\cite{glowinski1975sur,gabay1976dual}.
Observe that $\ccalL_{\beta}$ is strongly convex in the primal and auxiliary variables, yielding unique closed-form solutions for each sub-problem.
%

In a nutshell, \textit{our estimation method proceeds as follows}. We first use the trajectory $\ccalX$ to obtain the sample estimate $\tilde{\tenbQ}$ via \eqref{eq::emp_est_Q}. Then, we use $\tilde{\tenbQ}$ to run the optimization in \eqref{eq::auglag}, which yields the scaling vector $\hblambda$ and the factors $\{\hbQ_d,\hbQ_d'\}_{d=1}^D$. Upon substituting those into \eqref{eq::cpd_model}, we obtain our estimate $\hat{\tenbQ}$ for $q$. The last step is to use $\hat{\tenbQ}$ to obtain $\hat{\tenbR}$ and $\hat{\tenbP}$, the estimates for $\pi$ and $p$, via \eqref{eq::cpd_marginal} and \eqref{eq::cpd_joint}, respectively. 

Under appropriate choice of the parameter $\beta>0$, we provide the following guarantee that applying ADMM to~\eqref{eq::aux_prob} by cyclically minimizing~\eqref{eq::auglag} converges to a stationary solution of~\eqref{eq::opt_prob}~\cite{wang2019global}.

\vspace{-.2cm}

\begin{theorem}\label{thm:admm_conv}
    For a sufficiently large $\beta>0$ and any initial values $(\bbX^{(0)},\bbY^{(0)}, \bbW^{(0)})$ of the primal, auxiliary, and dual variables, solving~\eqref{eq::aux_prob} via ADMM generates a sequence of solutions $\{(\bbX^{(k)},\bbY^{(k)},\bbW^{(k)})\}_{k\geq 0}$ that converges to a stationary point of~\eqref{eq::opt_prob}.
\end{theorem}
\vspace{-.1cm}
The proof of Theorem~\ref{thm:admm_conv} can be found in the Appendix.
In brief, we have that $g$ in~\eqref{eq::aux_prob} obeys the assumptions on the objective function for Theorem 1 of~\cite{wang2019global}.
Moreover, the inclusion of the auxiliary variables $\bbpsi$, $\bbS$, and $\bbS'$ allows the remaining assumptions to be satisfied.
Thus, the result in~\cite[Theorem 1]{wang2019global} guarantees that ADMM for~\eqref{eq::aux_prob} converges to a stationary solution.
Then, since~\eqref{eq::aux_prob} is equivalent to~\eqref{eq::opt_prob}, the limit point is also a stationary solution of~\eqref{eq::opt_prob}.
Additionally, as the augmented Lagrangian $\ccalL_{\beta}$ in~\eqref{eq::auglag} is a Kurdyka-{\L}ojasiewicz function, ADMM will converge to a unique limit point starting from any initial point.
Thus, we solve the nonconvex problem in~\eqref{eq::opt_prob} by iterating over a set of convex sub-problems that provably converge to a stationary solution. 

\vspace{-.2cm}

\begin{remark}
    Under mild dimensionality conditions, the stationary distribution tensor $\tenbR$ has a unique decomposition~\cite{sidiropoulos2017tensor}.
    If $\tenbR$ further has rank $F$, then the decomposition in~\eqref{eq::cpd_marginal} is the unique CPD decomposition of $\tenbR$.
    Recall that by the ergodicity of our Markov chain, $\bbQ_d$ and $\bbQ'_d$ are interchangeable in~\eqref{eq::cpd_marginal}, so by the uniqueness of the decomposition, $\bbQ$ and $\bbQ'$ must be equivalent.
    In this case, we have a symmetric joint probability tensor $\tenbQ$, and the chain is thus reversible~\cite{aldous1982some}.
    Then, solving~\eqref{eq::opt_prob} only requires one set of factor matrices since $\bbQ=\bbQ'$.
    However, as we do not restrict our Markov models to be reversible, we proceed with a more general solution.
    In future work, we will explore the relationship between tensor rank and reversibility, both practically and theoretically.
\end{remark}

\vspace{-.1cm}


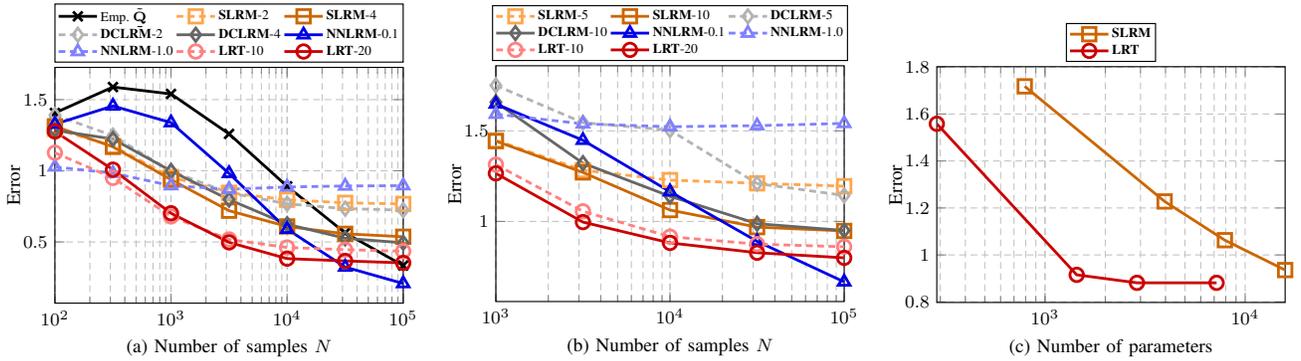
\begin{figure*}[t]
    \centering
    \hspace{0.7cm}
    \begin{minipage}[c]{.30\textwidth}
        \input{figs/varysamples_tensor}
    \end{minipage}
    \hspace{0.2cm}
    \begin{minipage}[c]{.30\textwidth}
        \input{figs/varysamples_ny}
    \end{minipage}
    \hspace{0.2cm}
    \begin{minipage}[c]{.30\textwidth}
        \vspace{.25cm}
        \input{figs/varyrank_ny}
    \end{minipage}
    \caption{
    (a) 
    Estimation error for synthetic low-rank transition tensor $\tenbP$ as the number of samples in the observed trajectory increase.
    (b)
    Estimation error for Markov chain based on real-world NYC taxi data as the number of samples in the observed trajectory increase.
    (c)
    Comparison of estimation error for \textbf{LRT} and \textbf{SLRM} as the number of parameters in each model increase.
    }
    \vspace{-.3cm}
\label{f:exps}
\end{figure*}

\section{Numerical evaluation}
\label{S:exp}

We demonstrate empirically the advantages of the low-rank tensor method proposed in~\eqref{eq::opt_prob}, denoted as ``\textbf{LRT}'', for estimating Markov chains.
By leveraging low-rankness in multi-dimensional state spaces, we can obtain the Markov chain with  fewer samples and parameters. 
We evaluate our approach in two settings: first, synthetic low-rank Markov chains, and second, a real-world Markov chain based on New York City taxi data~\cite{nyc_taxi_data}, a visualization of which is shown in Fig.~\ref{f:ny_data}a. 

We compare \textbf{LRT} against several existing low-rank matrix estimation techniques: (i)~\textbf{NNLRM}, matrix estimation with nuclear norm regularization~\cite{zhu2022learning}, (ii)~\textbf{DCLRM}, a relaxation of rank-constrained matrix estimation via difference of convex function (DC) programming~\cite{li2018estimation}, and (iii)~\textbf{SLRM}, matrix estimation with truncated singular values~\cite{zhang2019spectral}.
We let ``\textbf{LRT}-$M$'' denote \textbf{LRT} with tensor rank $M$, while ``\textbf{DCLRM}-$M$'' and ``\textbf{SLRM}-$M$'' denote estimation with matrix rank $M$.
Moreover, ``\textbf{NNLRM}-$c$'' represents \textbf{NNLRM} with $c$ as the nuclear norm penalty weight.
For matrix methods, we rearrange the tensors $\tenbP,\tenbQ\in [0,1]^{I_1 \times\dots I_D \times I_1 \times \dots I_D}$ as matrices of size $I \times I$.
Note that \textbf{LRT}-$M$ obtains a model of $(2\sum_{d} I_d +1)M$ parameters, that is, the number of entries in $\bblambda$, $\bbQ$, and $\bbQ'$.
Similarly, \textbf{SLRM}-$M$ estimates a matrix decomposition with $(2I+1)M$ parameters.
However, \textbf{NNLRM} and \textbf{DCLRM} are regularized matrix estimation methods and must estimate $I^2$ parameters, that is, all of the transition probabilities.
While \textbf{NNLRM} and \textbf{DCLRM} estimate $\tenbP$, the decomposition-based methods \textbf{LRT} and \textbf{SLRM} estimate $\tenbQ$, so we apply~\eqref{eq::cpd_marginal} and~\eqref{eq::cpd_joint} for our final approximations of $\tenbP$.
Further experimental details can be found at the associated GitHub repository~\cite{repository_github}.

\subsection{Synthetic low-rank Markov chains}

We first assess \textbf{LRT} and the matrix estimation baselines for estimating a synthetic Markov chain that follows a low-rank model.
We consider a multi-dimensional state space to illustrate the ability of our tensor model to capture low-rank dynamics while retaining the multi-dimensional structure of the state space.
In particular, the Markov chain is defined over a $D$-dimensional state space with $D=3$, where each dimension comprises $I_d=5$ elements.
The transition probabilities are represented by the tensor $\tenbP \in [0,1]^{(5 \times 5 \times 5) \times (5 \times 5 \times 5)}$ with rank $F = 10$.
We use the low-rank tensor chain to sample a trajectory $\ccalX$ of $N$ transitions, from which we construct the sample estimates~$\tilde{\tenbQ}$ via~\eqref{eq::emp_est_Q} and $\tilde{\tenbP}$ via~\eqref{eq::emp_est_P}. 
We employ \textbf{LRT} to obtain the estimate $\hat{\tenbQ}$, which is then used to estimate the transition probability tensor $\hat{\tenbP}$.
In tandem, we obtain transition probability estimates via \textbf{NNLRM}, \textbf{DCLRM}, and \textbf{SLRM}.
We generate trajectories for an increasing number of transitions from $N=10^2$ to $N=10^5$. 
For each value of $N$, we report the average normalized error $\| \hat{\tenbP} - \tenbP \|_1 / \|\tenbP\|_1$ over 10 trials in Fig.~\ref{f:exps}a. 

Our method \textbf{LRT} achieves competitive or superior performance for both the true tensor rank $F = 10$ and the more expressive $F = 20$.
Indeed, \textbf{LRT} outperforms \textbf{SLRM} and \textbf{DCLRM}, even though \textbf{SLRM}-2 and \textbf{SLRM}-4 possess more parameters than \textbf{LRT}-10 and \textbf{LRT}-20, respectively, while \textbf{DCLRM} has more parameters than both \textbf{LRT} and \textbf{SLRM}.
Observe that \textbf{NNLRM}-1.0 encourages matrix low-rankness, which does not comply with the tensor low-rankness of $\tenbP$, resulting in poor estimation even as the number of samples increases.
Conversely, \textbf{NNLRM}-0.1 has a milder low-rank penalty and prioritizes fitting to the empirical estimate $\tilde{\tenbP}$, obtaining an accurate estimate for a very large number of samples as $\tilde{\tenbP}$ approaches the true tensor $\tenbP$.
However, \textbf{LRT}-20 is a far simpler model with $(2\sum_d I_d + 1)F = 620 \ll I^2 = 5^6$ parameters, yet \textbf{LRT}-20 still rivals \textbf{NNLRM}-0.1. 

\subsection{New York City taxi data}

Next, we evaluate the performance of the \textbf{LRT} method using real-world data from the New York City taxi dataset, which has $D=2$ dimensions corresponding to the location and time period of each taxi stop.
Each transition corresponds to a taxi trip with the pick-up and drop-off locations and time periods.
We observe trips within Manhattan, where there are $I_0=66$ possible taxi stop locations and $I_1 = 6$ time periods, that is, partitions of a day into 4-hour segments, yielding $I=I_0I_1 = 396$ total states.
We let our ground truth Markov model $\tenbP$ be the estimate from~\eqref{eq::emp_est_P} using all $2.5$ million trips from January $2024$, depicted in Fig.~\ref{f:ny_data}b.
We compare \textbf{LRT} to the baselines as the number of samples increases from $N=10^3$ to $N=10^5$.

We again observe that \textbf{LRT} rivals or outperforms the baselines for tensor rank $F = 20$.
Our method \textbf{LRT} achieves lower error than \textbf{DCLRM} and \textbf{SLRM}, despite the larger number of parameters for \textbf{DCLRM} and \textbf{SLRM}-10.
Similarly to the synthetic case, \textbf{NNLRM}-1.0 with a stronger matrix low-rankness penalty yields suboptimal performance, while \textbf{NNLRM}-0.1 prioritizes fitting to the empirical estimate $\tilde{\tenbP}$, which is advantageous given a high number of samples.
However, given the natural multi-dimensionality of this scenario, \textbf{LRT} requires far fewer parameters to model the transition dynamics while rivaling \textbf{NNLRM}-0.1, despite \textbf{LRT}-20 being much less expressive.
Indeed, in low-sample regimes, the simpler model \textbf{LRT}-20 shows significant improvement over \textbf{NNLRM}-0.1.

We also demonstrate the value of modeling multi-dimensionality using tensors.
We present visualizations of the estimates of \textbf{LRT} and \textbf{SLRM} respectively in Figs.~\ref{f:ny_data}c and d, along with a comparison of their performance for a varying number of parameters given a fixed number of samples $N = 100, 000$ in Fig.~\ref{f:exps}c.
As both methods estimate decompositions, we can assess the value of considering tensors versus matrices for this real-world scenario. 
While both methods improve as the number of parameters increase, \textbf{LRT} requires far fewer parameters to outperform \textbf{SLRM}.
Even with $15, 840$ parameters for \textbf{SLRM}, our method \textbf{LRT} is able to improve the error for as low as $2, 880$ parameters.
Moreover, we observe that \textbf{SLRM} yields a block-wise estimate in Fig.~\ref{f:ny_data}c, but unlike \textbf{LRT} in Fig.~\ref{f:ny_data}d, it does not capture the block structure of the true tensor in Fig.~\ref{f:ny_data}b.
This indicates that not only can real-world data be naturally simplified by exploiting multi-dimensionality but also our tensor-based approach succeeds in capturing that multi-dimensional structure.


\section{Conclusion}
\label{S:conc}

In this work, we modeled Markov chains on multi-dimensional state spaces using low-rank tensors.
First, we showed that the CPD decomposition of our tensor model is equivalent to a naive Bayes representation of the joint transition probabilities.
This led to our approach to estimate a multi-dimensional Markov model from a sample trajectory as a low-rank tensor optimization problem.
Furthermore, we showed that solving the problem with ADMM converges to a stationary point.
Our synthetic and real-world simulations demonstrated our tensor model to be an efficient representation for multi-dimensional dynamical systems.
In particular, compared to traditional matrix-based algorithms, our low-rank tensor estimation achieved similar or superior performance with fewer parameters and less data.
In future work, we will investigate the effect of tensor rank on Markov models for multi-dimensional state spaces, such as the implications of rank for reversibility. 
Moreover, we can extend to other tensor decompositions for alternative notions of low-rankness.

\appendix{

For convenience, we let semicolons denote vertical concatenations, that is, for two vectors $\bbx \in \reals^m$ and $\bby \in \reals^n$, we have that $[\bbx; \bby] \in \reals^{m + n}$.
We also define the vectorization $\vect{\bbX}$ as the vertical concatenation of the columns of the matrix $\bbX$.
Moreover, let $J = \sum_{d=1}^D I_d$ be the sum of the state dimensions.

We first rewrite the optimization problem~\eqref{eq::aux_prob} in a vectorized form amenable for Theorem 1 of~\cite{wang2019global}.
We define $\bbq_d = \vect{\bbQ_d}$, $\bbq'_d = \vect{\bbQ'_d}$, $\bbs_d = \vect{\bbS_d}$, and $\bbs'_d = \vect{\bbS'_d}$ for every $d\in [D]$.
We collect the primal and auxiliary variables in the vectors $\bbx = [\bblambda; \bbq_1; \dots; \bbq_D; \bbq'_1; \dots; \bbq'_D]$ and $\bby = [\bbpsi; \bbs_1; \dots; \bbs_D; \bbs'_1; \dots; \bbs'_D]$, respectively, such that $\bbx,\bby \in \reals^{F(2J + 1)}$.
We further define the set of vectors $\bbx_0 = \bblambda$, $\bbx_d = \bbq_d$, and $\bbx_{D+d} = \bbq'_d$ for every $d\in[D]$.
Then, to rewrite the constraints of~\eqref{eq::aux_prob} in terms of $\bbx$ and $\bby$, we introduce the following matrices.
First, let $\bbE \in \reals^{(2DF+1) \times F(2J + 1)}$ be a block diagonal matrix
\alna{
    \bbE := \mathrm{blockdiag}\big(
        \bbone_F^\top,
        \bbI_F \otimes \bbone_{I_1}^\top,
        \dots,
        \bbI_F \otimes \bbone_{I_D}^\top,
    &\nonumber\\&
        \qquad \qquad \qquad \qquad \qquad \qquad
        \bbI_F \otimes \bbone_{I_1}^\top,
        \dots,
        \bbI_F \otimes \bbone_{I_D}^\top
    \big).
\nonumber
}
Then, we define
\alna{
    \bbA_0 :=
    [
        \bbI_F;
        \bbzero_{FI_1\times F};
        \dots;
        \bbzero_{FI_D\times F};
        \bbzero_{FI_1\times F};
    &\nonumber\\&
        \qquad \qquad
        \dots;
        \bbzero_{FI_D\times F};
        \bbzero_{2DF+1 \times F}
    ] \in \reals^{(F(2J+1) + 2DF+1) \times F}
\nonumber
}
associated with $\bbx_0$,
\alna{
    \bbA_1 :=
    [
        \bbzero_{F\times FI_1};
        \bbI_{FI_1};
        \dots;
        \bbzero_{FI_D\times FI_1};
        \bbzero_{FI_1\times FI_1};
    &\nonumber\\&
        \qquad \quad ~
        \dots;
        \bbzero_{FI_D\times FI_1};
        \bbzero_{2DF+1 \times FI_1}
    ] \in \reals^{(F(2J+1) + 2DF+1) \times FI_1}
\nonumber
}
with $\bbx_1$, and so on, yielding a collection of matrices $\{\bbA_i\}_{i=0}^{2D}$ such that $\sum_{i=0}^{2D} \bbA_i \bbx_i = [\bbx; \bbzero_{2DF+1}]$.
Then, we define $\bbA := [\bbA_0,\dots,\bbA_{2D}] = [\bbI_{F(2J + 1)}; \bbzero_{(2DF+1) \times F(2J + 1)}]$, $\bbB := [-\bbI_{F(2J+1)}; \bbE]$, and $\bbb := [\bbzero_{F(2J+1)}; \bbone_{2DF+1}]$.
We can then express the equality constraints in~\eqref{eq::aux_prob} as
\alna{
    \bbmat \bbx-\bby \\ \bbE\bby \ebmat
    =
    \sum_{i=0}^{2D} \bbA_i\bbx_i + \bbB\bby
    =
    \bbA\bbx + \bbB\bby = \bbb.
\nonumber
}

Finally, to rewrite the objective function in~\eqref{eq::aux_prob}, we define the mapping $\ccalT:\reals^{F(2J+1)} \rightarrow \reals^{ I_1\times\cdots\times I_D\times I_1 \times\cdots\times I_D }$ such that $\ccalT(\bbx) = \llbracket \bblambda, \bbQ,\bbQ' \rrbracket$.
The optimization problem~\eqref{eq::aux_prob} then is equivalent to
\alna{
    \min_{\bbx,\bby}~~
    \phi(\bbx) := \frac{1}{2} \| \tilde{\tenbQ} - \ccalT(\bbx) \|_F^2
    + \sum_{i=0}^{2D} \mbI\{ \bbx_i \geq \bbzero \}
    &\nonumber\\&
    \mathrm{s.t.}~~
    \bbA\bbx + \bbB\bby = \bbb
\label{prf:aux_prob}
}
with augmented Lagrangian
\alna{
    \ccalL_{\beta}(\bbx,\bby,\bbw)
    &~=~&
    \phi(\bbx)
    + \frac{\beta}{2} \norm{\bbA\bbx + \bbB\bby - \bbb}_2^2
    &\nonumber\\&
    &&
    + \inner{\bbw, \bbA\bbx + \bbB\bby - \bbb}.
\label{prf:aug_lag}
}
Observe that~\eqref{prf:aug_lag} is equivalent to $\ccalL_{\beta}$ in~\eqref{eq::auglag}, where $\bbw$ contains the vectorizations of the dual variables $\bbu$, $\bbU$, $\bbU'$, $v$, $\bbv$, and $\bbv'$.
In particular, $\bbw$ is defined as
\alna{
    \bbw :=
    [ \bbu; \vect{\bbU_1};  \dots; \vect{\bbU_D};  
    \vect{\bbU'_1};  
    &\nonumber\\&
    \qquad \qquad \qquad
    \dots; \vect{\bbU'_D}; 
    v; \bbv_1; \dots; \bbv_D; \bbv'_1; \dots; \bbv'_D ].
\nonumber
}
Indeed, the objective function values in matrix form in~\eqref{eq::aux_prob} and in vectorized form~\eqref{prf:aux_prob} are equivalent, and the squared $\ell_2$ norm and the inner product in $\ccalL_{\beta}$ as in~\eqref{prf:aug_lag} are separable and can be split into the terms in~\eqref{eq::auglag}.

With the optimization problem in the form~\eqref{prf:aux_prob}, we may now prove that the five necessary conditions hold for the conclusion of~\cite[Theorem 1]{wang2019global}.
In particular, we require that
\begin{itemize}[left= 13pt .. 22pt, noitemsep]
    \item[\textbf{AS1}] The objective function $\phi$ is coercive;
    \item[\textbf{AS2}] The solution is feasible, that is, $\Im{\bbA} \subseteq \Im{\bbB}$;
    \item[\textbf{AS3}] The problem~\eqref{prf:aux_prob} yields Lipschitz continuous sub-minimization paths for each primal vector $\bbx_i$;
    \item[\textbf{AS4}] For $g(\bbx) = \frac{1}{2} \| \tilde{\tenbQ} - \ccalT(\bbx) \|_F^2$ and $f_i(\bbx_i) = \mbI\{ \bbx_i \geq \bbzero \}$ for every $i\in\{0\} \cup [2D]$, $g$ is Lipschitz differentiable in any bounded set, $f_0$ is lower semi-continuous, and $f_i$ is restricted prox-regular for every $i\in[2D]$; and
    \item[\textbf{AS5}] The function $\phi$ is Lipschitz differentiable with respect to $\bby$.
\end{itemize}

It is trivial to see that \textbf{AS2} and \textbf{AS5} hold; $\Im{\bbA}$ is indeed a subspace of $\Im{\bbB}$, and $\phi$ is constant with respect to $\bby$.
Thus, we continue with \textbf{AS1} and show that $\phi$ is coercive.
Let $\phi(\bbx) < \infty$ be bounded for some $\bbx$.
By the presence of the indicator functions, each entry of $\bbx$ and thus of $\bblambda$, $\bbQ$, and $\bbQ'$ is bounded.
Therefore, the tensor $\ccalT(\bbx) = \llbracket \bblambda, \bbQ, \bbQ' \rrbracket$ has bounded entries, which are multilinear polynomials of entries of $\bblambda$, $\bbQ$, $\bbQ'$, or equivalently of $\bbx$.
Then, we must have that $\norm{\bbx}_2 < \infty$, so $\phi$ is coercive.

Assumption \textbf{AS3} requires that, for every $i\in\{0\}\cup[2D]$, the solution map $H_i(\bbz) = \{\argmin_{\bbx} \phi(\bbx)~\mathrm{s.t.}~\bbA_i\bbx_i = \bbz \}$ is Lipschitz continuous.
Since each $\bbA_i$ is full column rank, $H_i$ is a linear operator and thus Lipschitz continuous, satisfying \textbf{AS3}.

Next, we verify that \textbf{AS4} holds.
For a given $\bbx$ and a bounded open set $\ccalC$ containing $\bbx$, we choose a radius $r$ small enough such that the closed ball $\ccalN$ centered at $\bbx$ with radius $r$ is contained in $\ccalC$, that is, $\ccalN \subset \ccalC$.
Let $G(t) = \nabla_{\bbx} g( \bbx' + t (\bbx-\bbx') )$ for some $\bbx' \in \ccalN$.
By the mean value inequality, we have that 
\alna{
    \| \nabla_{\bbx} g(\bbx) - \nabla_{\bbx} g(\bbx') \|_2^2
    ~=~
    \| G(1) - G(0) \|_2
&\nonumber\\&
    \qquad\qquad\qquad
    ~\leq~
    \sup_{c \in (0,1)} \| G'(c) \|_2
&\nonumber\\&
    \qquad\qquad\qquad
    ~=~
    \sup_{c \in (0,1)} \| \nabla_{\bbx}^2 g(\bbx' + c (\bbx - \bbx')) (\bbx-\bbx') \|_2
&\nonumber\\&
    \qquad\qquad\qquad
    ~\leq~
    \sup_{c \in (0,1)} \| \nabla_{\bbx}^2 g(\bbx' + c (\bbx - \bbx') \|_F \cdot \| \bbx-\bbx' \|_2
&\nonumber\\&
    \qquad\qquad\qquad
    ~\leq~
    M \| \bbx - \bbx' \|_2,
\nonumber
}
where the last inequality holds since $\nabla_{\bbx}g$ is a multivariate polynomial and the ball $\ccalN$ is compact, so the mapping $\bbx \mapsto \| \nabla^2_{\bbx} g(\bbx) \|_F$ is continuous and attains a maximum $M$ over $\ccalN$ by the extreme value theorem.
Thus, $\nabla_{\bbx} g(\bbx)$ is locally Lipschitz, that is, $g(\bbx)$ is Lipschitz differentiable in any bounded set.
Moreover, indicator functions on convex sets are both lower semi-continuous and restricted prox-regular~\cite{wang2019global}.
Thus, we have that \textbf{AS4} holds.

We then have that all five conditions \textbf{AS1} to \textbf{AS5} hold for the result in~\cite[Theorem 1]{wang2019global}.
This states that ADMM initialized at any point $(\bbx^{(0)}, \bby^{(0)}, \bbw^{(0)})$ applied to~\eqref{prf:aux_prob} will yield a sequence of solutions $\{ (\bbx^{(k)}, \bby^{(k)}, \bbw^{(k)}) \}_{k\geq 0}$ that converges to a stationary point of~\eqref{prf:aux_prob} for a large enough $\beta > 0$. 
Moreover, observe that $g(\bbx)$ is a semi-algebraic function~\cite{attouch2010proximal} and thus $\ccalL_{\beta}$ a Kurdyka-{\L}ojasiewicz function. 
By Theorem 1 of~\cite{wang2019global}, we can further guarantee that the sequence will converge to the same stationary point regardless of the initial point $(\bbx^{(0)}, \bby^{(0)}, \bbw^{(0)})$.
Finally, as $\bbx$, $\bby$, and $\bbw$ contain vectorizations of the primal variables $\bblambda$, $\bbQ$, $\bbQ'$, the auxiliary variables $\bbpsi$, $\bbS$, $\bbS'$, and the dual variables $\bbu$, $\bbU$, $\bbU'$, $v$, $\bbv$, $\bbv'$, this implies that the convergence result holds when ADMM is applied to the matrix version of the problem~\eqref{eq::aux_prob}, which is equivalent to the original problem~\eqref{eq::opt_prob}.
$\hfill\blacksquare$
}


\bibliographystyle{ieeetr}
\bibliography{bibfile}

\end{document}

%% file: figs/varysamples_tensor.tex
\begin{tikzpicture}[baseline,scale=.85,trim axis left, trim axis right]


\pgfplotstableread{data/varysamp_ten_norml1.csv}\errtable

\definecolor{dblu}{RGB}{0,68,136} 
\definecolor{lblu}{RGB}{102,153,204} 
\definecolor{dyel}{RGB}{153,119,0} 
\definecolor{lyel}{RGB}{238,204,102} 
\definecolor{dred}{RGB}{153,68,85} 
\definecolor{lred}{RGB}{238,153,170} 
\definecolor{dgra}{RGB}{187,187,187} 
\definecolor{lgra}{RGB}{119,119,119} 

\begin{semilogxaxis}[
    xlabel={(a) Number of samples $N$},
    xmin=1e2,
    xmax=1e5,
    ylabel={Error},
    grid style=densely dashed,
    grid=both,
    legend style={
        at={(.5, 1.02)},
        anchor=south},
    legend columns=3,
    width=200,
    height=150,
    label style={font=\small},
    tick label style={font=\small}
    ]

    \addplot[black, mark=x, solid] table [x=Samples, y=Emp] {\errtable};
    
    \addplot[white!30!orange, mark=square, densely dashed] table [x=Samples, y=SLRM-2] {\errtable};
    \addplot[black!20!orange, mark=square, solid] table [x=Samples, y=SLRM-4] {\errtable};

    \addplot[white!70!black, mark=diamond, densely dashed] table [x=Samples, y=DCLRM-2] {\errtable};
    \addplot[white!40!black, mark=diamond, solid] table [x=Samples, y=DCLRM-4] {\errtable};

    \addplot[black!10!blue, mark=triangle, solid] table [x=Samples, y=NNLRM-0.1] {\errtable};
    \addplot[white!50!blue, mark=triangle, densely dashed] table [x=Samples, y=NNLRM-1.0] {\errtable};

    \addplot[white!50!red, mark=o, densely dashed] table [x=Samples, y=LRT-10] {\errtable};
    \addplot[black!20!red, mark=o, solid] table [x=Samples, y=LRT-20] {\errtable};

    \legend{Emp. $\tilde{\tenbQ}$, 
            \textbf{SLRM}-2, \textbf{SLRM}-4,
            \textbf{DCLRM}-2,\textbf{DCLRM}-4,
            \textbf{NNLRM}-0.1,\textbf{NNLRM}-1.0,
            \textbf{LRT}-10,\textbf{LRT}-20
            }

\end{semilogxaxis}

\end{tikzpicture}

%% file: figs/varysamples_ny.tex
\begin{tikzpicture}[baseline,scale=.85,trim axis left, trim axis right]


\pgfplotstableread{data/5_ny_sampling.csv}\errtable

\definecolor{dblu}{RGB}{0,68,136} 
\definecolor{lblu}{RGB}{102,153,204} 
\definecolor{dyel}{RGB}{153,119,0} 
\definecolor{lyel}{RGB}{238,204,102} 
\definecolor{dred}{RGB}{153,68,85} 
\definecolor{lred}{RGB}{238,153,170} 
\definecolor{dgra}{RGB}{187,187,187} 
\definecolor{lgra}{RGB}{119,119,119} 

\begin{semilogxaxis}[
    xlabel={(b) Number of samples $N$},
    xmin=1e3,
    xmax=1e5,
    ylabel={Error},
    grid style=densely dashed,
    grid=both,
    legend style={
        at={(.5, 1.02)},
        anchor=south},
    legend columns=3,
    width=200,
    height=150,
    label style={font=\small},
    tick label style={font=\small}
    ]

    
    \addplot[white!30!orange, mark=square, densely dashed] table [x=sampling, y=err_slrm_r5_P] {\errtable};
    \addplot[black!20!orange, mark=square, solid] table [x=sampling, y=err_slrm_r10_P] {\errtable};
    
    \addplot[white!70!black, mark=diamond, densely dashed] table [x=sampling, y=err_dclrm_5_P] {\errtable};
    \addplot[white!40!black, mark=diamond, solid] table [x=sampling, y=err_dclrm_10_P] {\errtable};
    
    \addplot[black!10!blue, mark=triangle, solid] table [x=sampling, y=err_nnlrm_01_P] {\errtable};
    \addplot[white!50!blue, mark=triangle, densely dashed] table [x=sampling, y=err_nnlrm_1_P] {\errtable};
    
    \addplot[white!50!red, mark=o, densely dashed] table [x=sampling, y=err_lrt_r10] {\errtable};
    \addplot[black!20!red, mark=o, solid] table [x=sampling, y=err_lrt_r20] {\errtable};

    \legend{
            \textbf{SLRM}-5, \textbf{SLRM}-10,
            \textbf{DCLRM}-5,\textbf{DCLRM}-10,
            \textbf{NNLRM}-0.1,\textbf{NNLRM}-1.0,
            \textbf{LRT}-10,\textbf{LRT}-20
            }

\end{semilogxaxis}

\end{tikzpicture}

%% file: figs/varyrank_ny.tex
\begin{tikzpicture}[baseline,scale=.85,trim axis left, trim axis right]


\pgfplotstableread{data/6_ny_params.csv}\errtable

\definecolor{dblu}{RGB}{0,68,136} 
\definecolor{lblu}{RGB}{102,153,204} 
\definecolor{dyel}{RGB}{153,119,0} 
\definecolor{lyel}{RGB}{238,204,102} 
\definecolor{dred}{RGB}{153,68,85} 
\definecolor{lred}{RGB}{238,153,170} 
\definecolor{dgra}{RGB}{187,187,187} 
\definecolor{lgra}{RGB}{119,119,119} 

\begin{semilogxaxis}[
    xlabel={(c) Number of parameters},
    xmin=288,
    xmax=15840,
    ylabel={Error},
    grid style=densely dashed,
    grid=both,
    legend style={
        at={(.5, 1.02)},
        anchor=south},
    legend columns=1,
    width=200,
    height=150,
    label style={font=\small},
    tick label style={font=\small}
    ]
    
    \addplot[black!20!orange, mark=square, solid] table [x=params_slrm, y=err_rank_slrm] {\errtable};
    \addplot[black!20!red, mark=o, solid] table [x=params_lrt, y=err_rank_lrt] {\errtable};

    \legend{\textbf{SLRM}, \textbf{LRT}}

\end{semilogxaxis}

\end{tikzpicture}